\documentclass[ALICE,manyauthors]{cernphprep}
\usepackage[comma,square,numbers,sort&compress]{natbib}

\usepackage{lineno}
\modulolinenumbers[5]
\usepackage{amssymb}  
\usepackage{xspace}
\usepackage{hyperref}
\usepackage{color}
\usepackage[T1]{fontenc}

\newcommand{\Snn}{\sqrt{s_{\mathrm{\rm NN}}}}

\newcommand{\Rz}{\ensuremath{\rho^{0}}\xspace}

\begin{document}

\begin{titlepage}
\PHyear{2021}       
\PHnumber{252}      
\PHdate{4 January}  

\title{First measurement of coherent  $\mathbf{\Rz}$ photoproduction in ultra-peripheral Xe--Xe collisions at  $\mathbf{\sqrt{s_{\rm {\scriptscriptstyle \mathbf{NN}}}} = 5.44}$  TeV}
\ShortTitle{Coherent $\mathbf{\Rz}$ photoproduction in ultra-peripheral Xe--Xe collisions}   

\Collaboration{ALICE Collaboration\thanks{See Appendix~\ref{app:collab} for the list of collaboration members}}
\ShortAuthor{ALICE Collaboration} 

\begin{abstract}
The first measurement of the coherent photoproduction of $\rho^{0}$ vector mesons in ultra-peripheral Xe--Xe collisions at  $\sqrt{s_{\mathrm{\rm NN}}}= 5.44$  TeV is presented. This result, together with previous HERA $\gamma$p data and $\gamma$--Pb measurements from ALICE, describes the atomic number ($A$) dependence of this process, which is particularly sensitive to nuclear shadowing effects and to the approach to the black-disc limit of QCD at a semi-hard scale. The cross section of the ${\rm Xe}+{\rm Xe} \to \Rz + {\rm Xe}+{\rm Xe}$ process, measured at midrapidity through the decay channel $\rho^{0}\to\pi^+\pi^-$, is found to be ${\rm d}\sigma/{\rm d}y=131.5\pm 5.6 {\rm (stat.)} ^{+17.5}_{-16.9} {\rm (syst.)}$ mb. The ratio of the continuum to resonant contributions for the production of pion pairs is also measured. In addition, the fraction of events accompanied by electromagnetic dissociation of either one or both colliding nuclei is reported. The dependence on $A$ of cross section for the coherent $\rho^{0}$ photoproduction at a centre-of-mass energy per nucleon of the $\gamma A$ system of $W_{\gamma A,n}= 65$ GeV is found to be consistent with a power-law behaviour $\sigma (\gamma A\rightarrow \rho^{0} A) \propto A^{\alpha}$ with a slope $\alpha = 0.96 \pm 0.02 {\rm (syst.)}$. This slope signals important shadowing effects, but it is still far from the behaviour expected  in the black-disc limit.

\end{abstract}
\end{titlepage}

\setcounter{page}{2} 


\section{Introduction
\label{sec:Intro}}

The Large Hadron Collider (LHC) is a source of photon-induced processes. The electromagnetic fields of the relativistic nuclei are strongly contracted allowing for their interpretation as a flux of quasi-real photons, which interact with the  nuclei travelling in the opposite direction. When the impact parameter of the collision is larger than the sum of the radii of the incoming nuclei, purely strong interactions are suppressed due to the short range of this force and photon-induced processes dominate. Interactions of this type are called ultra-peripheral collisions (UPCs)~\cite{Baltz:2007kq,Contreras:2015dqa,Klein:2019qfb}.

Among all the possible processes in UPC, the coherent production of vector mesons stands out due to the large associated cross sections and the cleanliness of its experimental signature: the  quasi-real photon interacts with the coherent QCD field of the other incoming particle to produce only a vector meson. Due to the coherence condition, the average transverse momentum of the vector meson, $\left<p_T\right>$, is related to the transverse size of the nucleus $R_A$ as $\left< p_T\right>\sim \hslash/R_A$~\cite{Baltz:2007kq}, yielding $\left< p_T\right>\sim 37$ (30) MeV/$c$ for a Xe (Pb) nucleus. A related process is the incoherent production where the photon interacts with a nucleon in the nucleus, which implies a larger average transverse momentum of the produced vector meson. In addition, secondary electromagnetic interactions of the colliding nuclei may excite one or both of them and upon de-excitation produce neutrons at beam rapidities~\cite{Baltz:2002pp}. This effect depends on the square of the electric charge of the nucleus, so it is expected to be substantially weaker for Xe than for Pb.

One of the photoproduction processes with the largest cross section is the production of a  $\Rz$ vector meson, which offers the opportunity to study the approach to the black-disc limit of QCD with a semi-hard scale~\cite{Frankfurt:2002wc}. 
This process has been extensively studied at the Relativistic Heavy Ion Collider (RHIC) in Au--Au,  and at the LHC in Pb--Pb UPC. Measurements at RHIC were performed by the STAR Collaboration at centre-of-mass energies per nucleon pair ($\Snn$) of 62.4 GeV~\cite{Agakishiev:2011me}, 130 GeV~\cite{Adler:2002sc}, and 200 GeV~\cite{Adamczyk:2017vfu}, while the studies at the LHC by  the ALICE Collaboration were carried out at  2.76 TeV~\cite{Adam:2015gsa} and 5.02 TeV~\cite{Acharya:2020sbc}. All measurements were performed at midrapidity. At the time of the first experimental results, the model predictions of the cross section varied substantially, with the model predicting the lowest cross section underestimating data by one standard deviation and the model predicting the largest cross section almost a factor two above data~\cite{Adam:2015gsa}. The availability of new and more precise data motivated an  improvement of the different theoretical approaches, which in general are within some 20\% to data~\cite{Acharya:2020sbc}. The situation, although better than a few years ago, still calls for more data to improve our understanding of this process. Furthermore, the coherent production of a  $\Rz$ vector meson off a nucleus allows for the study of shadowing, the experimental fact that the nuclear structure functions are suppressed compared to the superposition of those of their constituent nucleons~\cite{Armesto:2006ph}. This phenomenon is expected to depend on the atomic number $A$ of the nucleus so measurements for different values of $A$ offer another tool to test our understanding of shadowing at high energies and semi-hard scales.

In collisions of two heavy ions with atomic number $A$, either nucleus can be a source of photons, which results in two contributions to the cross section. At midrapidity both contributions are the same, but at forward rapidities one corresponds to a high-energy photon while the other to a low-energy photon. If one could disentangle both contributions it would be possible to study in the same experiment the energy dependence of the process, allowing for the study of the energy dependence of the underlying QCD dynamics. Two techniques have been put forward to this end~\cite{Contreras:2016pkc,Guzey:2013jaa}. The first one makes use of UPCs and peripheral collisions. The second, described in Ref.~\cite{Guzey:2013jaa}, proposes to classify the measured events depending on the presence of the beam-rapidity neutrons mentioned earlier, and use these different cross sections in each class to disentangle the low and high energy contributions. Measurements at midrapidity are ideal to test this proposal because both contributions are the same, so the measured cross sections can be unambiguously compared with models predicting the neutron emission probability. ALICE can measure beam-rapidity neutrons at both sides of the nominal interaction point. 
The sides are called A and C, with the latter one hosting the ALICE muon spectrometer~\cite{Aamodt:2008zz}.
ALICE has previously published~\cite{Acharya:2020sbc} the cross section for the coherent production of $\Rz$ vector mesons in Pb--Pb UPC for events with no beam-rapidity neutrons (0n0n, where the first 0n refers to the A-side and the second to the C-side), with one or more neutrons on one side only (0nXn+Xn0n), or on both sides (XnXn). 
The comparison of the corresponding cross section fractions with calculations of the emission of beam-rapidity neutrons based on the STARlight~\cite{Klein:1999qj,Klein:2016yzr} and $\textbf{n$\mathbf{_O^O}$n}$~\cite{Broz:2019kpl} models suggests that the method works, but it is important to test it further, for example with the large data sets expected from the LHC Run 3 and 4~\cite{Citron:2018lsq}.

 In this letter, the first measurement of the coherent photoproduction of $\Rz$ vector mesons in  Xe--Xe UPCs at  $\Snn= 5.44$  TeV is presented. The cross section of this process is measured at midrapidity through the decay channel $\Rz\to\pi^+\pi^-$. The ratio of the continuum to resonant contributions for the production of pion pairs is also measured. In addition, the fraction of events in the 0n0n, 0nXn+Xn0n, and XnXn classes are given. Finally, using data from HERA and from Pb--Pb UPC collisions measured by ALICE, the $A$ dependence of the cross section is studied at a  centre-of-mass energy per nucleon of the $\gamma A$ system of $W_{\gamma A,n}=65$ GeV.

\section{Experimental set-up 
\label{sec:Setup}}
During a 6-hour pilot run in October 2017,  the LHC collided xenon nuclei for the first time. These collisions took place at $\Snn = 5.44$ TeV.  A complete description of the ALICE detector and its performance can be found in Ref.~\cite{Aamodt:2008zz,Abelev:2014ffa}; here, just a brief description of the systems involved in the measurement is given.

The decay products of the $\Rz$ vector meson are measured in the central-barrel region of ALICE with the ITS and TPC detectors. The ALICE Inner Tracking System (ITS)~\cite{Aamodt:2010aa} is made of six layers of silicon sensors. Each layer has a cylindrical geometry concentric around the beam line. Three different technologies are used: pixel, drift and strip sensors. Each technology is used in two consecutive layers. All six layers are used for tracking in this analysis.
The Time-Projection Chamber (TPC)~\cite{Alme:2010ke} surrounds the ITS. It is a large cylindrical gas detector with a central membrane at high voltage and two readout planes, composed of multiwire proportional chambers, at the end caps. It is the main tracking detector and it also offers particle identification through the measurement of ionisation energy loss.  The TPC and ITS cover  a pseudorapidity interval $|\eta|<0.9$ and the full azimuth; they are situated inside a large solenoid magnet which in Xe--Xe collisions provided a B = 0.2~T field.

The neutrons at beam rapidity are measured by two neutron zero-degree calorimeters, ZNA and ZNC, located at $\pm112.5$~m from the nominal interaction point along the beam line and covering the pseudorapidity range $|\eta|>8.8$~\cite{Abelev:2014ffa}.  The energy resolution for single neutrons is around 20\% which allows measuring of events with either zero or a few neutrons at beam rapidities.

A dedicated UPC trigger was utilised to collect the data. The trigger used the Silicon Pixel Detector (SPD), the Time-of-Flight detector (TOF) and the V0 detectors. The SPD forms the two innermost layers of the ITS, covering pseudorapidity ranges $|\eta|<2$ and $|\eta|<1.4$, respectively. The SPD has about $10^7$ pixels which are read out by 400 (800) chips in the inner (outer) layer. Each of the readout chips fires a trigger if at least one of its pixels has a signal. TOF surrounds the TPC and matches its pseudorapidity coverage. The TOF is a large cylindrical barrel of multigap resistive plate chambers  with some $1.5\times10^5$ readout channels arranged in 1608 pads that are capable of triggering~\cite{Akindinov:2009zzc}. The V0~\cite{Abbas:2013taa} is a set of two arrays made of 32 scintillator cells each. The arrays cover the pseudorapidity ranges $-3.7<\eta<-1.7$ and $2.8<\eta<5.1$, respectively. The time resolution of the V0 is better than 500~ps and provides a trigger if it registers a signal. The trigger requires at least two hits in the inner and in the outer layer of the SPD, at least two pads fired in TOF and no signal in the V0.

The determination of the luminosity is based on reference trigger counts in the V0 detector, while the reference trigger cross section is the product of the hadronic inelastic cross section, determined based on a Glauber model~\cite{Loizides:2017ack}, and a trigger efficiency factor. In detail, the actual trigger signal used as a reference for the luminosity determination requires a signal in the V0 detector with a total amplitude above a specific threshold, optimised for the  rejection of both beam-induced and electromagnetic (EM) background. In order to evaluate the reference trigger efficiency (ratio of the trigger cross section to the total hadronic cross section), the V0 signal amplitude distribution in minimum-bias events (collected with the trigger defined in~\cite{Acharya:2018hhy}) is fitted with a model which combines the Glauber model (for the centrality) and a negative binomial distribution (for particle production), as described in~\cite{Abelev:2013qoq},~\cite{Acharya:2018hhy}, and~\cite{Adam_2016}. The fit is performed in the 0--90\% centrality (where the minimum-bias trigger is fully efficient for hadronic interactions and fully inefficient for EM  interactions) and the distribution is then extrapolated to \mbox{0--100\%} to get the total integral. The trigger efficiency is finally determined as the number of events firing the reference trigger divided by the extrapolated integral of the minimum-bias spectrum. The trigger efficiency thus determined is $68.81\pm0.01$\,(stat.)\,\%.
In the Glauber model the following values for xenon are used: $ A = 129$, the radius of the nuclear-charge distribution  $r = (5.36 \pm 0.1)$~fm, a skin depth of $(0.59\pm 0.07)$ fm, and a deformation parameter $\beta^2 = 0.18 \pm 0.02$~\cite{Acharya:2018hhy}. The integrated luminosity used in this analysis is $(279.5\pm29.9)$~\textrm{mb}$^{-1}$, where the quoted uncertainty is systematic and is described later.

\section{Analysis procedure
\label{sec:Ana}}

\begin{figure}[t!]
\centering
\includegraphics[width=0.48\textwidth]{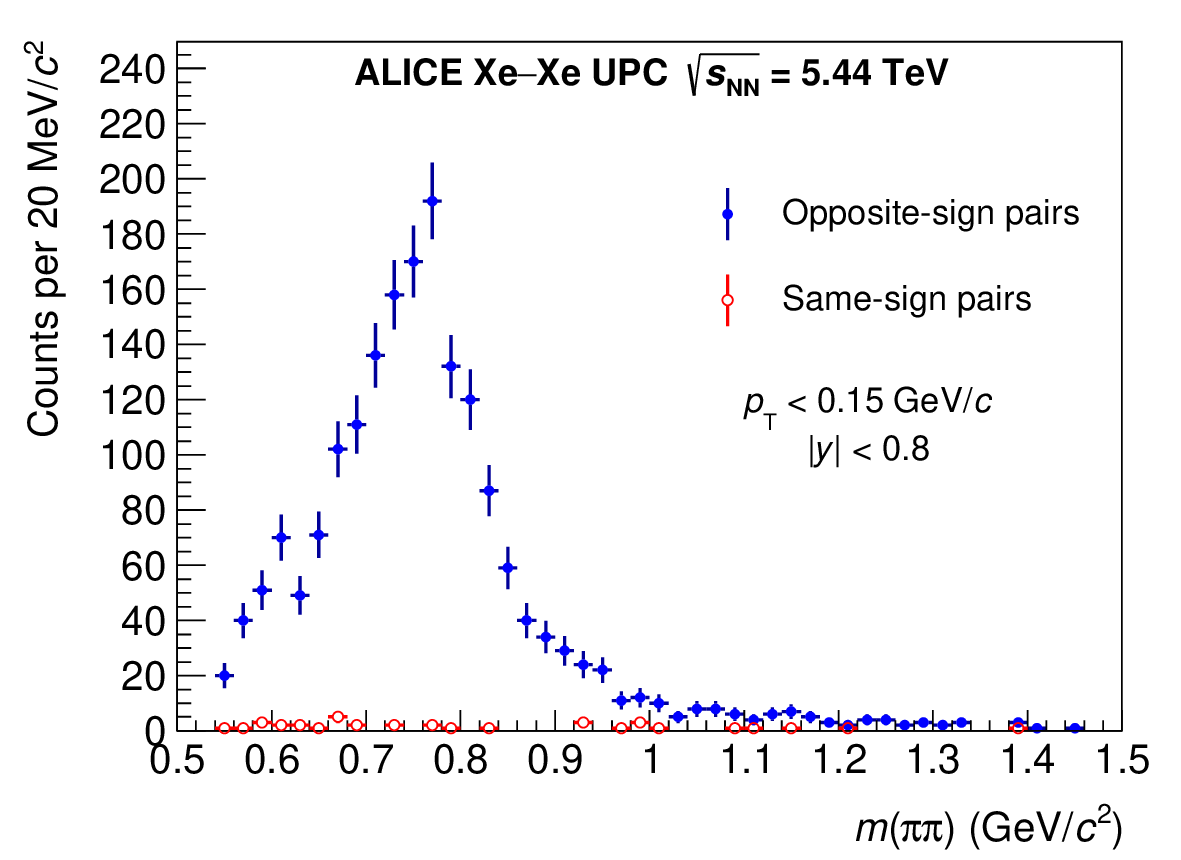}
\includegraphics[width=0.48\textwidth]{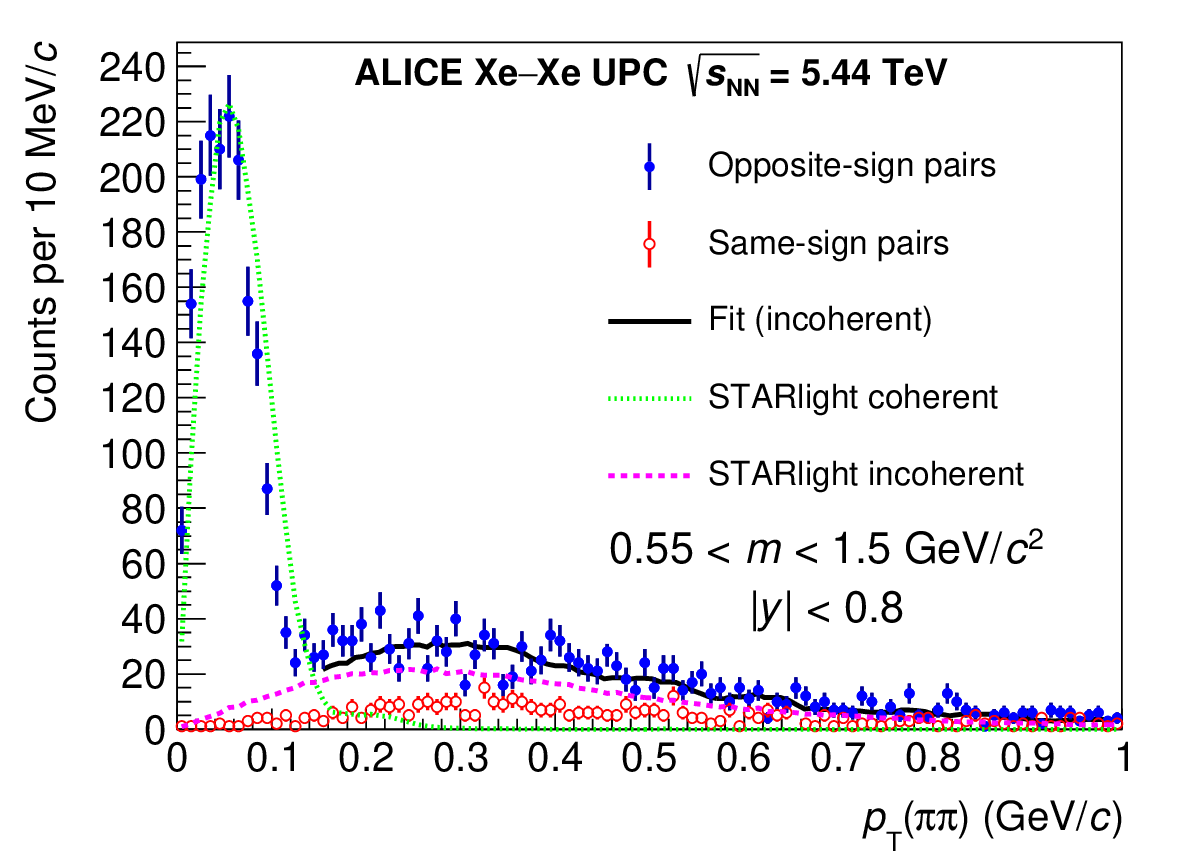}
\caption{\label{fig:kine} (Colour online) Uncorrected invariant mass (left) and transverse momentum (right) distribution of dipion candidates. Also shown are track pairs that have the same electric charge and fulfil all other requirements. The STARlight templates for coherent and incoherent production---shown in the right panel with the green and magenta lines, respectively---are normalised to the corresponding luminosity of data. An example of a fit to obtain the incoherent contribution (see text for details) is also shown (black line).
}
\end{figure}
\subsection{Event and track selection}

Events are selected for the  analysis if  ($i$) the trigger described above is active, ($ii$) there are no signals in the V0 detectors as determined by an offline selection, and ($iii$) they have exactly two tracks fulfilling the requirements listed below. 

Offline, a more refined algorithm to quantify the V0 timing signal is used, consisting of a larger time window. For this reason, this analysis requires the V0 offline reconstruction for selecting events.

The tracks are required to have contributions from both the ITS and the TPC. Both layers of the SPD have to have a signal associated both to the track and to a SPD trigger signal, the tracks should also have at least 50 (out of 159) space points reconstructed by the TPC. 
Both tracks are required to be fully within the acceptance of the detector ($|\eta_{\rm trk}| < 0.8$).  They have to originate from a  primary vertex whose coordinate along the beam line fulfils $|z_{\rm trk}|<10$~cm, and their associated  electric charge should be of opposite sign. Particle identification of a track is determined by the number of standard deviations ($n_{\sigma}$) by which the energy loss measurement deviates from the pion hypothesis. The quadratic sum of $n_{\sigma_{1,2}}$ for the $\pi^+$ and $\pi^-$ candidates has to be less than five squared ($n^2_{\sigma_1}+n^2_{\sigma_2} < 5^2$). Finally, the transverse momentum of the pion pair has to be less than 0.15 GeV/$c$. After these selections, 1827 events remain in the data sample.

Figure~\ref{fig:kine} shows the distributions of the invariant mass of the pion pairs as well as their transverse momentum. In addition, the figure also shows the corresponding distribution of an alternate data sample obtained by applying all the criteria described above except that the electric charge associated to both tracks has to be of the same sign. The mass distribution shows a clear signal of a $\Rz$ vector meson over a very small background represented by the same-sign distribution. The transverse momentum distribution shows a pronounced peak at values of a few tens of MeV/$c$ as expected by coherent production accompanied by a  tail towards larger momenta produced by incoherent production and a small remaining  background, e.g. from peripheral hadronic collisions of the two incoming nuclei. Transverse momentum distributions of MC events, after applying on them the same selection criteria as real data (Sec.~\ref{sec:SignalExtraction}), are also shown in the figure to support the interpretation of the distribution being dominated by coherent production at low and incoherent interactions at larger transverse momenta. An example of a fit, described in more detail in Sec.~\ref{sec:SysUnc}, to obtain the incoherent contribution is also shown.

The presence of one or more neutrons at beam rapidity is determined by using the timing capabilities of the ZNA and ZNC, which allow for the selection of events with a signal within $\pm2$~ns from the time expected for neutrons produced in the interaction. 

\subsection{Signal extraction
\label{sec:SignalExtraction}}

A Monte Carlo (MC) sample of pion pairs from continuum  and $\Rz$ resonant production, generated with the STARlight program~\cite{Klein:2016yzr}, is used to extract mass-dependent efficiency correction factors to account for the acceptance  and the efficiency of the detector and the selection criteria. All events in this sample are passed by a detailed simulation of the ALICE apparatus and subjected to the same analysis procedure as in data. The  correction factor at each mass of the pion pair is used to correct the mass distribution shown in Fig.~\ref{fig:kine}. The correction factor increases from about 0.025 at 550 MeV/$c^2$ to 0.055 at  1.1 GeV/$c^2$. The number of $\Rz$ vector mesons is extracted from the corrected invariant mass distribution normalised by the luminosity of the sample and after applying other corrections, described below, to take into account pile-up and the contribution from incoherent events.

The corrected mass distribution is fitted to a model describing the resonant and continuum production of pion pairs according to the S\"oding prescription~\cite{Soding:1965nh} and a term $M$ that takes into account  the contribution of the $\gamma\gamma\to\mu^+\mu^-$ process:
\begin{equation}
	\frac{{\rm d}^2\sigma}{{\rm d}m\,{\rm d}y} = |A \times BW_{\rho}+B|^2+M.
	\label{eq:Soeding}
\end{equation}
Here $A$ is the normalisation factor of the $\Rz$ Breit--Wigner ($BW_{\rho}$) function, and $B$ is the non-resonant amplitude. The relativistic Breit--Wigner function of the $\Rz$ vector meson is
\begin{equation}
	BW_{\rho} = \frac{\sqrt{m \times m_{\rho^0} \times \Gamma(m)}}
                  {m^2- m^2_{\rho^0} + im_{\rho^0} \times \Gamma(m)},
	\label{SoedingBW}
\end{equation}
where $m_{\Rz}$ is the pole mass of the $\Rz$ vector meson. The mass-dependent width $\Gamma(m)$ is given by
\begin{equation}
	\Gamma(m) = \Gamma(m_{\rho^0}) \times \frac{m_{\rho^0}}{m} \times
         \left( \frac{m^2-4m^2_{\pi}}{m^2_{\rho^0}-m^2_{\pi}} \right)^{3/2},
	\label{SoedingWidth}
\end{equation}
with $\Gamma(m_{\Rz})$ the width of the $\Rz$ vector meson and $m_{\pi}$  the mass of the pion~\cite{Jackson:1964zd}. The shape of $\gamma\gamma\to\mu^+\mu^-$ process $M$ is taken from STARlight and passed  through the same selection procedure as data. The fitted parameters are $A$, $B$ and $M$, while the mass and width of the $\Rz$ were fixed to the PDG values~\cite{PDG:2020}.

\begin{figure}[t!]
\centering
\includegraphics[width=0.80\textwidth]{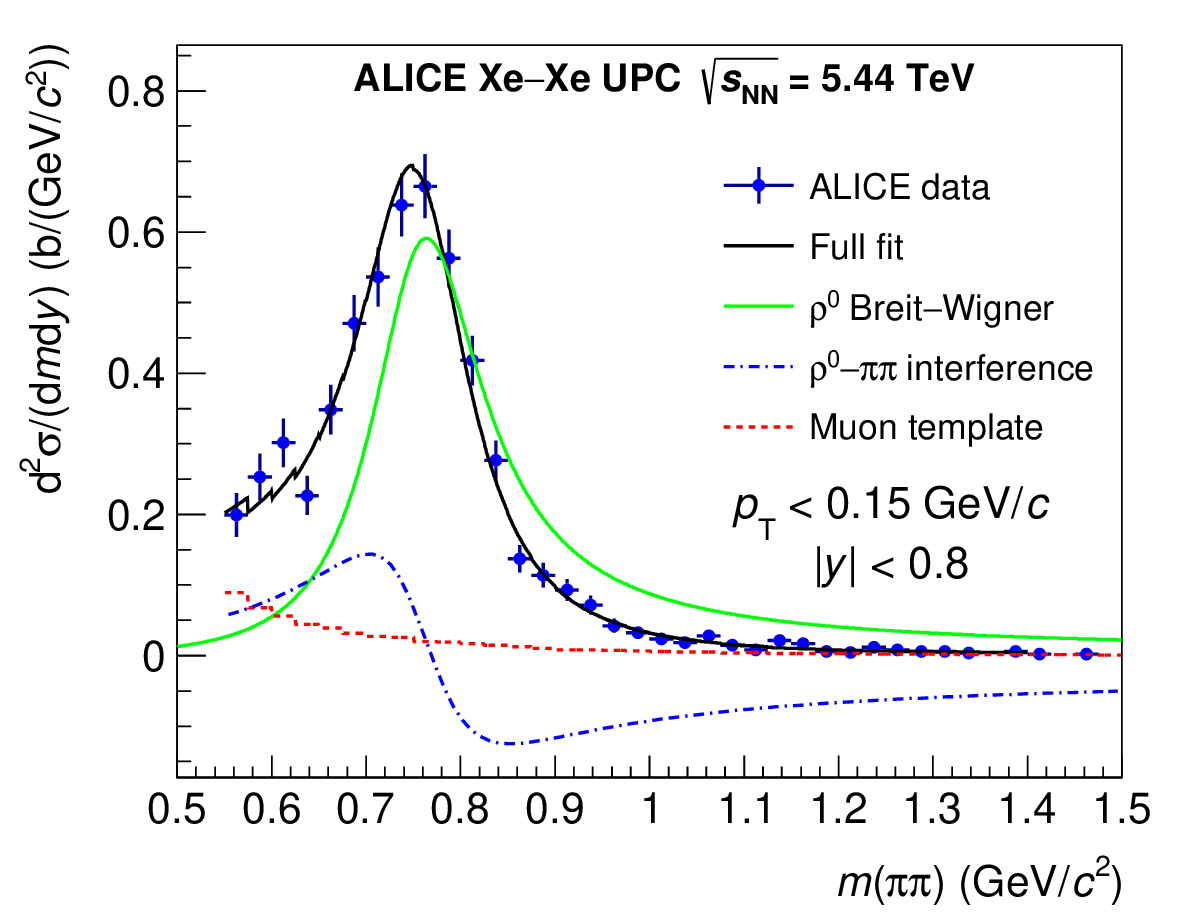}
\caption{\label{fig:fit} (Colour online)  Invariant mass distribution of pion pairs with the different components of the fit represented by lines. The number of $\Rz$ candidates is integrated over the resonant Breit--Wigner part (green, full line), the interference term between $A$ and $B$ of Eq.~(\ref{eq:Soeding}) is shown by the dash-dotted blue line and the muon template $M$ (red, dashed line) is taken from  STARlight. See text for more details. 
}
\end{figure}

An example fit of data with this model is shown in Fig.~\ref{fig:fit}, where a clear resonance structure is seen in data and decomposed by the fit 
into  a small background contribution ($M$), the contribution of the continuum production of pion pairs ($B$), the $\Rz$ signal ($BW_{\rho}$) and the interference term from squaring the amplitude (see Eq.~\ref{eq:Soeding}).

\subsection{Remaining corrections and systematic uncertainties
\label{sec:SysUnc}}

The contributions to the systematic uncertainty are listed in Table~\ref{tab:sys} and discussed one by one in the following paragraphs. The total uncertainty is obtained as the quadratic sum of the various contributions.

\begin{table}[t!]
\centering
\caption{Summary of the systematic uncertainties for the measured cross section. See text for details.}
\begin{tabular}{lr}
\hline
 Source & Uncertainty   \\
 \hline
Variations to the fit procedure & $\pm 2.5$\% \\
Ross--Stodolsky fit model & $+3.5\%$  \\ 
Acceptance and efficiency & $\pm 0.5$\%  \\ 
Track selection & $\pm 3.0$\% \\
Track ITS--TPC matching & $\pm 4.0$\% \\
SPD trigger-to-track matching & $\pm 2.0$\%\\
TOF trigger efficiencies & $\pm 2.8$\%\\
Vertex selection & $\pm 1.5$\% \\
Incoherent contribution& $\pm 2.0$\% \\ 
Pile-up & $\pm 1.0$\%\\
Muon background ($\gamma\gamma\to\mu^+\mu^-$) & $^{+({\textrm 0.5})}_{-({\textrm 0.2})}$\% \\ 
Electromagnetic dissociation & $\pm 0.2$\%\\ 
Luminosity & $\pm10.7$\%  \\
\hline
Total & $^{+({\textrm 13.3})}_{-({\textrm 12.8})}$ \% \\
\hline
\end{tabular}
\label {tab:sys}
\end{table}

The signal extraction procedure is repeated many times with slightly different settings where the lower and upper limit of the fit range as well as the bin width are varied. The variations are within 0.55 to 0.65 GeV$/c^2$, 0.9 to 1.4 GeV$/c^2$, and 10 to 50 MeV$/c^2$, respectively. The average value of  the pole mass for the $\Rz$ vector meson as well as its width are found to agree with the value reported by the Particle Data Group~\cite{PDG:2020}. As the precision of our data sample is limited, the values for the  pole mass  and width of  the $\Rz$ vector meson are then fixed to the values measured in the $\Rz$ photoproduction process~\cite{PDG:2020} and the signal extraction procedure is repeated. The standard deviation of the distribution of extracted number of $\Rz$ vector mesons from all the different fits is considered as a systematic uncertainty, while the mean is taken as the signal. The mean of all statistical uncertainties is taken as the statistical uncertainty. Fits are performed using a log-likelihood as well as a $\chi^2$ approach; both producing the same results. The systematic uncertainty of the cross section from those variations amounts to 2.5\%.

The Ross--Stodolsky prescription~\cite{Ross:1965qa} is used as an alternative model to fit the resonance and continuum contribution,  which results in 
a yield systematically higher by 3.5\%. Pure MC studies, where signal is generated with a S\"oding function and fitted with a Ross--Stodolsky model, and vice versa, show a similar behaviour. As the underlying distribution is not known, this difference is taken as a systematic uncertainty. 

Two MC samples are used to account for the acceptance and efficiency correction. One simulates only the Breit--Wigner distribution for a pure $\Rz$ signal and the other includes the effect of the pion-pair continuum. The full variation of 0.5\% on the cross section obtained by using these two samples is considered a systematic uncertainty.

All the analysis steps are repeated by varying the tracking selection criteria within reasonable values. In particular, a test is performed where events with tracks in some parts of the detector, known to have reduced performance, are rejected. The full variation of the results amounts to 3\% and it is taken as a systematic uncertainty. Likewise, the uncertainty when matching track segments in the ITS to their counterparts in the TPC is studied by varying the track selection and investigating the dependence of the matching on the track kinematics and the general event characteristics, 
in each case comparing the results to those from a detailed simulation of the detector. This uncertainty amounts to 4\%.

The two pion tracks are matched to signals in the SPD. In real data it could be that the SPD has other signals, e.g. from noise or soft electron-positron production. The performance of the matching algorithm is checked by comparing the results of applying it in data and in MC, where these extra effects are not present. A discrepancy of 2.0\% is found and assigned as a systematic uncertainty. The uncertainty on the trigger efficiency of TOF is obtained by comparing the acceptance-times-efficiency correction obtained from MC under different assumptions and assigning the full 2.8\% difference.

There is a small discrepancy between data  and the MC description of the coordinate of the interaction vertex along the beam line  for collisions happening  $\pm 10$ cm and more beyond the nominal interaction point.  The full difference in the cross section when retaining (or not) events beyond $\pm 10$ cm is found to be $\pm 1.5$\% and assigned as a systematic uncertainty.

The contribution from incoherent production of $\Rz$ vector mesons for the region $p_{\rm T}<0.15$~GeV/$c$ is determined by fitting the corresponding template from STARlight, accounting for the same-sign contribution taken from data (see Fig.~\ref{fig:kine}), to the transverse momentum distribution in a range from 0.15 to 1.0 GeV/$c$ and extrapolating to the region covered by the measurement. The fit is repeated many times varying the fit ranges, within the stated interval, and the bin widths. An example of such a fit is shown in Fig.~\ref{fig:kine}. The  contribution from the incoherent production to the yield is given by the mean over the results from these fits;  it amounts to 10.2\% and it is subtracted  from the cross section. The standard deviation of all fits is taken as a systematic uncertainty ($\pm 2.0$\%). It is worth noting that the contribution from incoherent production varies across different neutron classes. As the data sample of events with neutrons at beam rapidities is small, the incoherent contribution is estimated for the full 0nXn+Xn0n+XnXn sample. 
The incoherent background amounts to $(6.1 \pm 3.0 {\rm (syst.)})$\% for the 0n0n and $(35.8 \pm 4.3{\rm (syst.)})$\% for the 0nXn+Xn0n+XnXn event classes, respectively. This is taken into account when computing the fractions of the cross section in each class reported below.

The V0 veto could be invalidated if this detector shows a signal which originates from a separate interaction, an effect called pile-up. Electromagnetic $e^+e^-$ pair production is the main source of these  signals.
The probability for pile-up is  obtained from an unbiased sample triggered by the timing of expected bunch crossings at the interaction point surrounded by the ALICE detector. This probability is used, assuming a Poisson process, to correct for the events lost due to pile-up. The correction factor is $0.89\pm0.01$; the statistical uncertainty from this procedure is taken as systematic uncertainty ($\pm 1.0$\%). 

The statistical uncertainty of the $\gamma\gamma\to {\rm e}^+{\rm e}^-$ cross section  in our previous measurement~\cite{Kryshen:2019jnz} is around 10\% and within this precision it agrees with the prediction from STARlight. Changing the normalisation of the $\gamma\gamma\to\mu^+\mu^-$ template in the fit by $\pm10$\%, produces a $-0.2$\% and $+0.5$\% systematic uncertainty on the extracted $\Rz$ cross section. 

Electromagnetic dissociation producing the beam-rapidity neutrons is accompanied on occasion by other charged particles~\cite{Pshenichnov:1999hw}. These charged particles, if they hit the V0, may cause the event to be lost. The probability for this to happen is estimated to be $(1.7\pm0.2)$\% using the unbiased sample just mentioned. The statistical precision of this procedure is taken as systematic uncertainty ($\pm 0.2$\%). 

The uncertainty on the luminosity is determined by computing the prediction of the Glauber model varying each parameter within their reported uncertainty. This is the dominant source of uncertainty in the measurement and amounts to 10.7\%. 

The extraction of fractions of the cross section in the 0n0n, 0nXn+Xn0n, and XnXn classes is also affected by pile-up in the ZNA and ZNC and by the efficiency of these calorimeters to detected neutrons. The pile-up probabilities are $(0.47\pm0.02)$\% and $(0.44\pm0.02)$\%, while the efficiencies are $0.91\pm0.01$ and $0.92\pm0.02$ for the ZNA and ZNC, respectively. The uncertainties in these numbers, along with the uncertainty on the subtraction of the incoherent contribution, are taken into account to obtain the uncertainty on the fractions quoted below.

\section{Cross section results
\label{sec:Results}}

The  cross section for the coherent photoproduction of $\Rz$ vector mesons in ultra-peripheral Xe--Xe collisions at  $\Snn= 5.44$  TeV measured at midrapidity is
\begin{equation}
\frac{{\rm d}\sigma}{{\rm d}y}=131.5\pm 5.6\ {\rm (stat.)} ^{+17.5}_{-16.9}\ {\rm (syst.)} \ {\rm mb.}
\end{equation}

\begin{figure}[!t]
\centering
\includegraphics[width=0.80\textwidth]{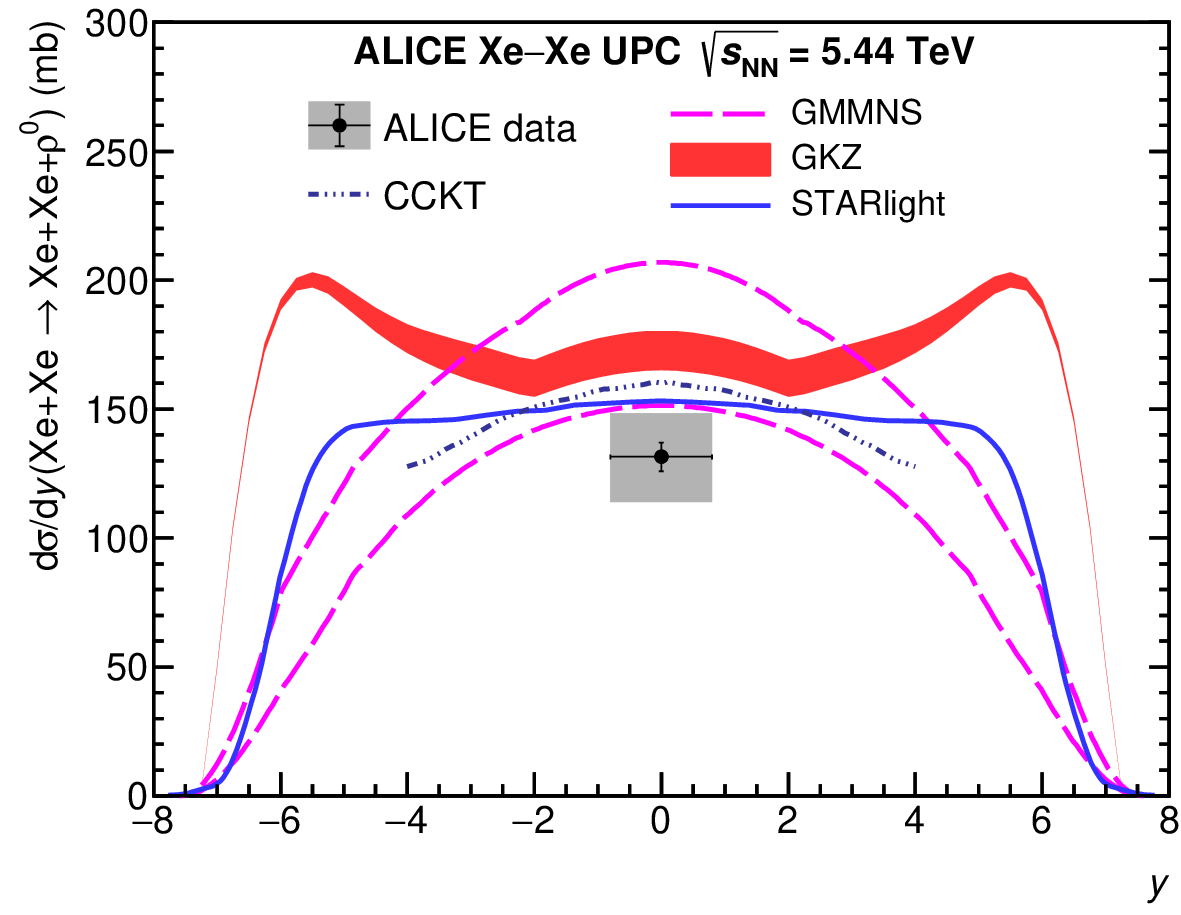}
\caption{\label{fig:crossSection} (Colour online) 
Cross section for the coherent photoproduction of $\Rz$ vector mesons in Xe--Xe UPC. The lines show the predictions of the different models described in the text.
}
\end{figure}

Figure~\ref{fig:crossSection} shows the measured cross section and compares it with the prediction of the following models. STARlight~\cite{Klein:2016yzr}, which is based on ($i$) a phenomenological description of existing data on exclusive production of $\Rz$ vector mesons off protons, ($ii$) the optical theorem, and ($iii$) a Glauber-like eikonal formalism which neglects the elastic part of the elementary $\Rz$--nucleon cross section. The prediction by Guzey, Kryshen and Zhalov (GKZ)~\cite{Guzey:2018bay} relies on a modified vector dominance model, where  hadronic fluctuations of the photon are taken into account according to the Gribov--Glauber model of nuclear shadowing; the band shows the variation on the predictions when varying the parameters of the model. The model by Gon\c{c}alves et al.\ (GMMNS)~\cite{Goncalves:2017wgg} uses the colour dipole approach with amplitudes obtained from the IIM model~\cite{Iancu:2003ge} coupled to a Glauber prescription to go from the nucleon to the nuclear case. The two lines shown for  GMMNS  bracket the changes in their predictions when varying different ingredients of their model~\cite{Goncalves:2017wgg}. Finally, the model from  Cepila et al.\ (CCKT)~\cite{Cepila:2016uku,Cepila:2018zky} also uses the colour dipole approach, but in this case the structure of the nucleon in the transverse plane is described by so-called hot spots, regions of high gluonic density, whose number increases with increasing energy~\cite{Cepila:2016uku}; nuclear effects are implemented along the ideas of the Glauber model proposed in Ref.~\cite{Armesto:2002ny}. At midrapidity, all  models are relatively close to one another and overestimate the data. The lower band of GMMNS as well as the STARlight and CCKT predictions are slightly more than one standard deviation above the data. Only the upper band of GMMNS is disfavoured by more than three standard deviations.

The ratio of non-resonant to resonant pion production, see Eq.~(\ref{eq:Soeding}), is measured to be $|B/A| = 0.58\pm 0.04\ ({\rm stat.)}\pm 0.03\ ({\rm syst.)}$ $({\rm GeV}/c^2)^{-\frac{1}{2}}$. The main uncertainty comes from the correction for acceptance and efficiency, closely followed by variations from the signal extraction procedure. This value is consistent with those obtained in Pb--Pb UPC at $\Snn=2.76$~TeV~\cite{Adam:2015gsa} and $\Snn=5.02$~TeV~\cite{Acharya:2020sbc}, namely $|B/A| = 0.50\pm 0.04\ ({\rm stat.)} \pm ^{0.10}_{0.04}\ ({\rm syst.)}$ $({\rm GeV}/c^2)^{-\frac{1}{2}}$ and $|B/A| = 0.57\pm 0.01\ ({\rm stat.)}\pm 0.02\ ({\rm syst.)}$ $({\rm GeV}/c^2)^{-\frac{1}{2}}$, respectively. The corresponding ratio in coherent Au--Au UPC measured by STAR at $\Snn = 200$~GeV is  $0.79\pm0.01\ ({\rm stat.)}\pm0.08\ ({\rm syst.)}$ $({\rm GeV}/c^2)^{-\frac{1}{2}}$~\cite{Adamczyk:2017vfu}. The CMS Collaboration measured $0.50\pm0.06\ ({\rm stat.)}$ $({\rm GeV}/c^2)^{-\frac{1}{2}}$ in p--Pb UPC at $\Snn=5.02$~TeV~\cite{Sirunyan:2019nog} for $|t|<0.5$~GeV$^2$. The ZEUS Collaboration, using a sample of positron--proton collisions at a centre-of-mass energy of 300 GeV,  reports $0.67\pm0.02\ ({\rm stat.)}\pm0.04\ ({\rm syst.)}$ $({\rm GeV}/c^2)^{-\frac{1}{2}}$ for their full analysed sample, and $\approx 0.8$ $({\rm GeV}/c^2)^{-\frac{1}{2}}$ for $t$ values similar to those of  coherent $\Rz$ production in Pb--Pb UPC~\cite{Breitweg:1997ed}.

The fraction of the cross section in each one of the  classes defined by the presence or absence of beam-rapidity neutrons  is shown in Table~\ref{tab:frac}, where the measurement is also compared with the prediction from the $\textbf{n$\mathbf{_O^O}$n}$ MC~\cite{Broz:2019kpl}. 
This program generates neutrons emitted due to the electromagnetic dissociation (EMD)  of two interacting nuclei. It is based on photon fluxes computed in the semi-classical approximation, and on all existing  data on EMD complemented by phenomenological extrapolations where data is not available. It can easily be interfaced to theoretical predictions of coherent vector meson production. 
The agreement of the model with data is at the level of one standard deviation; this,  as well as the satisfactory description of the corresponding cross sections observed in Pb--Pb UPC at $\Snn=5.02$~TeV~\cite{Acharya:2020sbc}, suggests that the emission of neutrons at beam-rapidity is well understood for the coherent photoproduction of $\Rz$ vector mesons off nuclei with such different atomic mass number as Pb and Xe.

\begin{table}[t!]
\centering
\caption{Fraction of the cross section in each one of the classes defined by the presence or absence of beam-rapidity neutrons compared with the predictions from the $\textbf{n$\mathbf{_O^O}$n}$ model~\cite{Broz:2019kpl}. The first uncertainty is statistical, the second comes from the variations in the ZNA and ZNC pile-up factors and efficiencies, while the third  comes from the variation in the number of events which is dominated by the subtraction of incoherent contribution. The use of $\pm$ or $\mp$ reflects the correlation between the classes. See text for details. }
{\renewcommand{\arraystretch}{1.5}
\begin{tabular}{lrr} 
\hline
Class &  Measured fraction & $\textbf{n$\mathbf{_O^O}$n}$ prediction\\
\hline
0n0n  &($90.46\pm0.70\pm0.17\mp0.68$)\% & 92.4\%\\
0nXn+Xn0n &  ($8.48\pm0.66\mp0.13\pm0.64$)\% & 6.9\% \\
XnXn &   ($1.07\pm0.25\mp0.04\pm0.07$)\% & 0.7\%\\
\hline
\end{tabular}
}
\label {tab:frac}
\end{table}

\begin{figure}[!t]
\centering
\includegraphics[width=0.80\textwidth]{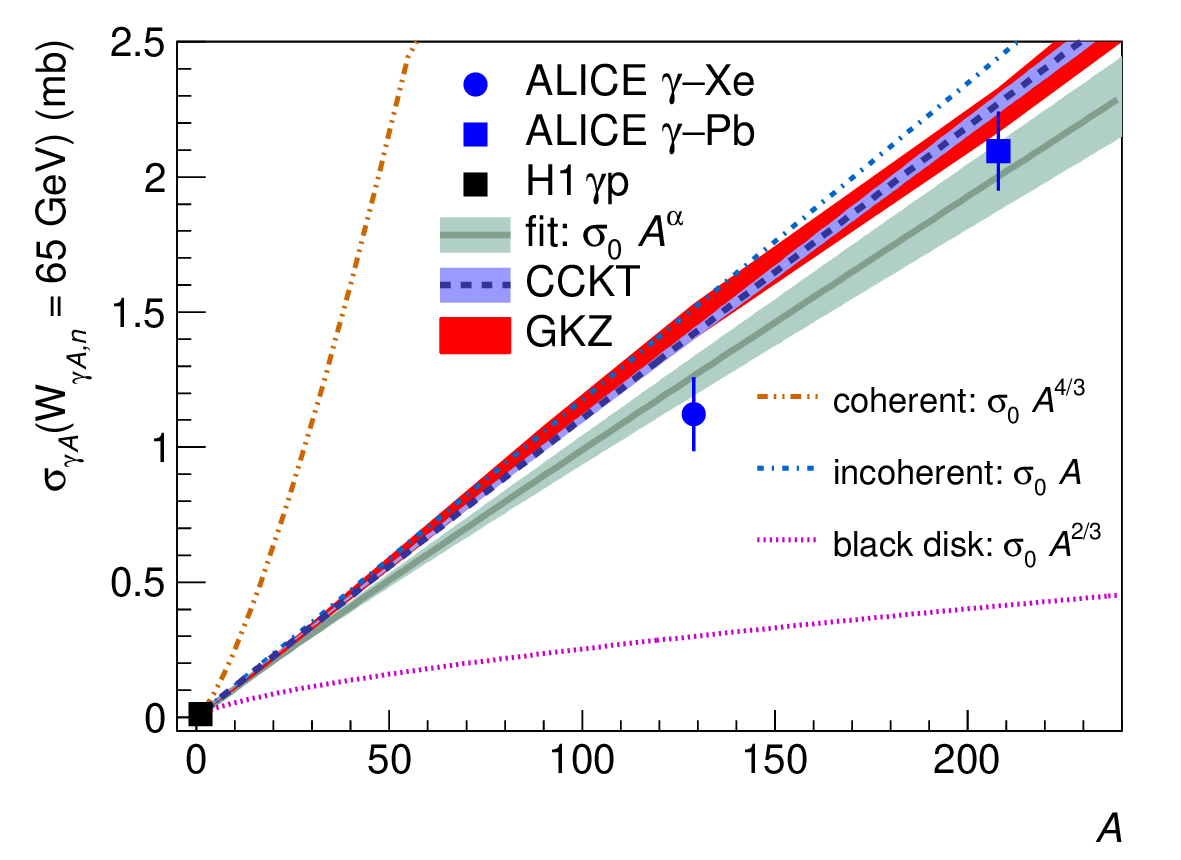}
\caption{\label{fig:xs_A} $A$ dependence of the  $\gamma A$ cross section for the coherent production of a $\Rz$ meson and the corresponding power-law fit shown as a band.  The data are from this analysis and from~\cite{Acharya:2020sbc,H1:2020lzc}. The general expectations for three extreme cases are represented by the dashed, dotted-dashed, and dotted lines, respectively. The red band corresponds to the GKZ predictions when varying the parameters of the model.  A power-law fit to the CCKT model is shown by the blue band. See text for details.}
\end{figure}

The measurements of the UPC  cross section for coherent production of $\Rz$ vector mesons at midrapdity for Pb--Pb~\cite{Acharya:2020sbc} and for Xe--Xe have been converted into a $\gamma A$ measurement by dividing the cross sections by two times the corresponding photon fluxes of 58.6 (Xe) and 128.1 (Pb). These numbers are obtained following the prescription detailed in Sec. II of Ref.~\cite{Contreras:2016pkc}. A flux uncertainty of 2\% is considered, which is uncorrelated between both nuclei, because it mainly originates in the knowledge of the nuclear geometries; specifically, it is obtained by a variation of the nuclear charge radius and skin thickness within their uncertainties as parameters of the Wood-Saxon distribution. The uncertainties coming from the Ross--Stodolsky fit model and from the ITS-TPC matching are  correlated between the Xe and Pb results. The midrapidity photon--nucleus centre-of-mass energy per nucleon is given by $W_{\gamma A,n}^{2}=m\Snn$ (with $m$ the mass of the vector meson), so it is
slightly different in both systems (62 GeV in Pb--Pb and 65 GeV in Xe--Xe); as the $\gamma$--Pb cross section  is expected to change around 1\% between these two values, well within the experimental uncertainties, both  measurements are taken as having  $W_{\gamma A,n} = 65$ GeV.

The dependence of these cross sections on $A$ is fitted by a power-law model, $\sigma_{\gamma A}(A) = \sigma_0A^{\alpha}$, using also the cross section measured by H1 at this energy~\cite{H1:2020lzc}: $(11.8\pm0.9(\rm{syst.}))$ $\mu$b. The value reported by H1 is consistent with the corresponding cross section found by the ZEUS~\cite{Breitweg:1997ed} and CMS~\cite{Sirunyan:2019nog} collaborations.
The fit is shown in Fig.~\ref{fig:xs_A}. It has a $\chi^2=1.48$ (for one degree of freedom). The parameters from the fit  when using only uncorrelated uncertainties are $\sigma_0 = 0.0117 \pm 0.0009$ mb and $\alpha =  0.963 \pm 0.019$. The correlation between them is ${-0.78}$. Varying the flux by $\pm2$\% produces a change in the exponent $\alpha $ of $0.005$. Varying the cross sections by the correlated uncertainties from the fit model and the ITS--TPC matching does not modify the $\sigma_0$ parameter and causes a change in the exponent $\alpha$ of  $+0.006$ and $\pm0.007$, respectively.

The fit is compared with three generic expectations having different dependence on $A$ resulting on slopes $\alpha$ of 4/3, 1, and 2/3 for full coherence disregarding any other dynamical effect,  for a total incoherent behaviour, and for the black-disc limit, respectively.
The slope found in data is significantly different from 4/3 signalling important shadowing effects. The closeness of data to a slope of 1 does not imply incoherent behaviour; it is just a coincidence produced by the large shadowing suppression. The black-disc limit seems to be quite distant at this energy of $W_{\gamma A}=65$ GeV. 

Fitting to the same functional form the predictions of the Gribov--Glauber approach (GKZ~\cite{Guzey:2016piu,Guzey:2018bay}) and of the colour dipole model with subnucleon degrees of freedom (CCKT~\cite{Cepila:2018zky,Krelina:2019gee}) yields slopes of $0.985\pm0.007$ and $0.984\pm0.003$, respectively, where in both cases the parameter $\sigma_0$ has been fixed to the corresponding prediction for the $\gamma$p cross section. Both slopes are in good agreement with that found in data. This was to be expected given that both approaches give a reasonable description of the different available data. 

\section{Summary
\label{sec:Summary}}

The cross section for the coherent photoproduction of $\Rz$ vector mesons in Xe--Xe UPC at $\Snn=5.44$ TeV has been measured and compared with existing models of this process. The theoretical predictions slightly overestimate the measurement. The ratio of the continuum-to-resonant contributions for the production of pion pairs is also measured and found to agree with previous measurements in Pb--Pb UPC. The fraction of events accompanied by electromagnetic dissociation of either one or both colliding nuclei is reported and compared with the predictions of the $\textbf{n$\mathbf{_O^O}$n}$ model. The fair agreement between data and predictions suggest that this process is well understood within the current experimental uncertainties and can be used as a tool to disentangle the different $\gamma A$ contributions to the UPC cross sections.

The dependence on $A$ of the cross section for  the coherent $\rho^{0}$ photoproduction at a centre-of-mass energy per nucleon of the $\gamma A$ system of 65 GeV is found to be consistent with a power-law behaviour with a slope of $0.96\pm0.02$. This exponent is substantially smaller than what is expected from a purely coherent process, taking into account the geometry, but disregarding any dynamic effect. A fair description of Pb--Pb and Xe--Xe data is obtained in models based on hadronic degrees of freedom in the Gribov--Glauber approach (GKZ) as well as in partonic-level models (CCKT).
In this context, the $A$ dependence of the cross section is a strong indicator that QCD effects are important and relatively well modelled. 


\newenvironment{acknowledgement}{\relax}{\relax}
\begin{acknowledgement}
\section*{Acknowledgements}


The ALICE Collaboration would like to thank all its engineers and technicians for their invaluable contributions to the construction of the experiment and the CERN accelerator teams for the outstanding performance of the LHC complex.
The ALICE Collaboration gratefully acknowledges the resources and support provided by all Grid centres and the Worldwide LHC Computing Grid (WLCG) collaboration.
The ALICE Collaboration acknowledges the following funding agencies for their support in building and running the ALICE detector:
A. I. Alikhanyan National Science Laboratory (Yerevan Physics Institute) Foundation (ANSL), State Committee of Science and World Federation of Scientists (WFS), Armenia;
Austrian Academy of Sciences, Austrian Science Fund (FWF): [M 2467-N36] and Nationalstiftung f\"{u}r Forschung, Technologie und Entwicklung, Austria;
Ministry of Communications and High Technologies, National Nuclear Research Center, Azerbaijan;
Conselho Nacional de Desenvolvimento Cient\'{\i}fico e Tecnol\'{o}gico (CNPq), Financiadora de Estudos e Projetos (Finep), Funda\c{c}\~{a}o de Amparo \`{a} Pesquisa do Estado de S\~{a}o Paulo (FAPESP) and Universidade Federal do Rio Grande do Sul (UFRGS), Brazil;
Ministry of Education of China (MOEC) , Ministry of Science \& Technology of China (MSTC) and National Natural Science Foundation of China (NSFC), China;
Ministry of Science and Education and Croatian Science Foundation, Croatia;
Centro de Aplicaciones Tecnol\'{o}gicas y Desarrollo Nuclear (CEADEN), Cubaenerg\'{\i}a, Cuba;
%
Ministry of Education, Youth and Sports of the Czech Republic, Czech Republic; Czech Science Foundation;
The Danish Council for Independent Research | Natural Sciences, the VILLUM FONDEN and Danish National Research Foundation (DNRF), Denmark;
Helsinki Institute of Physics (HIP), Finland;
Commissariat \`{a} l'Energie Atomique (CEA) and Institut National de Physique Nucl\'{e}aire et de Physique des Particules (IN2P3) and Centre National de la Recherche Scientifique (CNRS), France;
Bundesministerium f\"{u}r Bildung und Forschung (BMBF) and GSI Helmholtzzentrum f\"{u}r Schwerionenforschung GmbH, Germany;
General Secretariat for Research and Technology, Ministry of Education, Research and Religions, Greece;
National Research, Development and Innovation Office, Hungary;
Department of Atomic Energy Government of India (DAE), Department of Science and Technology, Government of India (DST), University Grants Commission, Government of India (UGC) and Council of Scientific and Industrial Research (CSIR), India;
Indonesian Institute of Science, Indonesia;
Istituto Nazionale di Fisica Nucleare (INFN), Italy;
Institute for Innovative Science and Technology , Nagasaki Institute of Applied Science (IIST), Japanese Ministry of Education, Culture, Sports, Science and Technology (MEXT) and Japan Society for the Promotion of Science (JSPS) KAKENHI, Japan;
Consejo Nacional de Ciencia (CONACYT) y Tecnolog\'{i}a, through Fondo de Cooperaci\'{o}n Internacional en Ciencia y Tecnolog\'{i}a (FONCICYT) and Direcci\'{o}n General de Asuntos del Personal Academico (DGAPA), Mexico;
Nederlandse Organisatie voor Wetenschappelijk Onderzoek (NWO), Netherlands;
The Research Council of Norway, Norway;
Commission on Science and Technology for Sustainable Development in the South (COMSATS), Pakistan;
Pontificia Universidad Cat\'{o}lica del Per\'{u}, Peru;
Ministry of Science and Higher Education, National Science Centre and WUT ID-UB, Poland;
Korea Institute of Science and Technology Information and National Research Foundation of Korea (NRF), Republic of Korea;
Ministry of Education and Scientific Research, Institute of Atomic Physics and Ministry of Research and Innovation and Institute of Atomic Physics, Romania;
Joint Institute for Nuclear Research (JINR), Ministry of Education and Science of the Russian Federation, National Research Centre Kurchatov Institute, Russian Science Foundation and Russian Foundation for Basic Research, Russia;
Ministry of Education, Science, Research and Sport of the Slovak Republic, Slovakia;
National Research Foundation of South Africa, South Africa;
Swedish Research Council (VR) and Knut \& Alice Wallenberg Foundation (KAW), Sweden;
European Organization for Nuclear Research, Switzerland;
Suranaree University of Technology (SUT), National Science and Technology Development Agency (NSDTA) and Office of the Higher Education Commission under NRU project of Thailand, Thailand;
Turkish Atomic Energy Agency (TAEK), Turkey;
National Academy of  Sciences of Ukraine, Ukraine;
Science and Technology Facilities Council (STFC), United Kingdom;
National Science Foundation of the United States of America (NSF) and United States Department of Energy, Office of Nuclear Physics (DOE NP), United States of America.
\end{acknowledgement}

\bibliographystyle{utphys}   
\bibliography{bibliography}

\providecommand{\href}[2]{#2}\begingroup\raggedright\begin{thebibliography}{10}

\bibitem{Baltz:2007kq}
A.~J. Baltz {\em et~al.}, ``{The Physics of Ultraperipheral Collisions at the
  LHC}'', \href{http://dx.doi.org/10.1016/j.physrep.2007.12.001}{{\em Phys.
  Rept.} {\bfseries 458} (2008) 1--171},
\href{http://arxiv.org/abs/0706.3356}{{\ttfamily arXiv:0706.3356 [nucl-ex]}}.

\bibitem{Contreras:2015dqa}
J.~G. Contreras and J.~D. Tapia~Takaki, ``{Ultra-peripheral heavy-ion
  collisions at the LHC}'',
\href{http://dx.doi.org/10.1142/S0217751X15420129}{{\em Int. J. Mod. Phys.}
  {\bfseries A30} (2015) 1542012}.

\bibitem{Klein:2019qfb}
S.~R. Klein and H.~Mäntysaari, ``{Imaging the nucleus with high-energy
  photons}'', \href{http://dx.doi.org/10.1038/s42254-019-0107-6}{{\em Nature
  Rev. Phys.} {\bfseries 1} no.~11, (2019) 662--674},
  \href{http://arxiv.org/abs/1910.10858}{{\ttfamily arXiv:1910.10858
  [hep-ex]}}.

\bibitem{Baltz:2002pp}
A.~J. Baltz, S.~R. Klein, and J.~Nystrand, ``{Coherent vector meson
  photoproduction with nuclear breakup in relativistic heavy ion collisions}'',
  \href{http://dx.doi.org/10.1103/PhysRevLett.89.012301}{{\em Phys. Rev. Lett.}
  {\bfseries 89} (2002) 012301},
\href{http://arxiv.org/abs/nucl-th/0205031}{{\ttfamily arXiv:nucl-th/0205031
  [nucl-th]}}.

\bibitem{Frankfurt:2002wc}
L.~Frankfurt, M.~Strikman, and M.~Zhalov, ``{Signals for black body limit in
  coherent ultraperipheral heavy ion collisions}'',
  \href{http://dx.doi.org/10.1016/S0370-2693(02)01882-8}{{\em Phys. Lett. B}
  {\bfseries 537} (2002) 51--61},
  \href{http://arxiv.org/abs/hep-ph/0204175}{{\ttfamily arXiv:hep-ph/0204175}}.

\bibitem{Agakishiev:2011me}
{\bfseries STAR} Collaboration, G.~Agakishiev {\em et~al.}, ``{$\rho^{0}$
  Photoproduction in AuAu Collisions at $\sqrt{s_{NN}}$=62.4 GeV with STAR}'',
  \href{http://dx.doi.org/10.1103/PhysRevC.85.014910}{{\em Phys. Rev.}
  {\bfseries C85} (2012) 014910},
\href{http://arxiv.org/abs/1107.4630}{{\ttfamily arXiv:1107.4630 [nucl-ex]}}.

\bibitem{Adler:2002sc}
{\bfseries STAR} Collaboration, C.~Adler {\em et~al.}, ``{Coherent $\rho^0$
  production in ultraperipheral heavy ion collisions}'',
  \href{http://dx.doi.org/10.1103/PhysRevLett.89.272302}{{\em Phys. Rev. Lett.}
  {\bfseries 89} (2002) 272302},
\href{http://arxiv.org/abs/nucl-ex/0206004}{{\ttfamily arXiv:nucl-ex/0206004
  [nucl-ex]}}.

\bibitem{Adamczyk:2017vfu}
{\bfseries STAR} Collaboration, L.~Adamczyk {\em et~al.}, ``{Coherent
  diffractive photoproduction of $\rho^0$ mesons on gold nuclei at 200
  GeV/nucleon-pair at the Relativistic Heavy Ion Collider}'',
  \href{http://dx.doi.org/10.1103/PhysRevC.96.054904}{{\em Phys. Rev.}
  {\bfseries C96} (2017) 054904},
\href{http://arxiv.org/abs/1702.07705}{{\ttfamily arXiv:1702.07705 [nucl-ex]}}.

\bibitem{Adam:2015gsa}
{\bfseries ALICE} Collaboration, J.~Adam {\em et~al.}, ``{Coherent $\rho^{0}$
  photoproduction in ultra-peripheral Pb-Pb collisions at $
  \sqrt{s_{\mathrm{NN}}}=2.76 $ TeV}'',
  \href{http://dx.doi.org/10.1007/JHEP09(2015)095}{{\em JHEP} {\bfseries 09}
  (2015) 095},
\href{http://arxiv.org/abs/1503.09177}{{\ttfamily arXiv:1503.09177 [nucl-ex]}}.

\bibitem{Acharya:2020sbc}
{\bfseries ALICE} Collaboration, S.~Acharya {\em et~al.}, ``{Coherent
  photoproduction of $\rho^{0}$ vector mesons in ultra-peripheral Pb-Pb
  collisions at $ \sqrt{{\mathrm{s}}_{\mathrm{NN}}} $ = 5.02 TeV}'',
  \href{http://dx.doi.org/10.1007/JHEP06(2020)035}{{\em JHEP} {\bfseries 06}
  (2020) 035}, \href{http://arxiv.org/abs/2002.10897}{{\ttfamily
  arXiv:2002.10897 [nucl-ex]}}.

\bibitem{Armesto:2006ph}
N.~Armesto, ``{Nuclear shadowing}'',
  \href{http://dx.doi.org/10.1088/0954-3899/32/11/R01}{{\em J. Phys. G}
  {\bfseries 32} (2006) R367--R394},
  \href{http://arxiv.org/abs/hep-ph/0604108}{{\ttfamily arXiv:hep-ph/0604108}}.

\bibitem{Contreras:2016pkc}
J.~G. Contreras, ``{Gluon shadowing at small $x$ from coherent
  $\mathrm{J/}\psi$ photoproduction data at energies available at the CERN
  Large Hadron Collider}'',
  \href{http://dx.doi.org/10.1103/PhysRevC.96.015203}{{\em Phys. Rev. C}
  {\bfseries 96} no.~1, (2017) 015203},
  \href{http://arxiv.org/abs/1610.03350}{{\ttfamily arXiv:1610.03350
  [nucl-ex]}}.

\bibitem{Guzey:2013jaa}
V.~Guzey, M.~Strikman, and M.~Zhalov, ``{Disentangling coherent and incoherent
  quasielastic $J/\psi$ photoproduction on nuclei by neutron tagging in
  ultraperipheral ion collisions at the LHC}'',
  \href{http://dx.doi.org/10.1140/epjc/s10052-014-2942-z}{{\em Eur. Phys. J.}
  {\bfseries C74} (2014) 2942},
\href{http://arxiv.org/abs/1312.6486}{{\ttfamily arXiv:1312.6486 [hep-ph]}}.

\bibitem{Aamodt:2008zz}
{\bfseries ALICE} Collaboration, K.~Aamodt {\em et~al.}, ``{The ALICE
  experiment at the CERN LHC}'',
\href{http://dx.doi.org/10.1088/1748-0221/3/08/S08002}{{\em JINST} {\bfseries
  3} (2008) S08002}.

\bibitem{Klein:1999qj}
S.~R. Klein and J.~Nystrand, ``{Exclusive vector meson production in
  relativistic heavy ion collisions}'',
  \href{http://dx.doi.org/10.1103/PhysRevC.60.014903}{{\em Phys. Rev.}
  {\bfseries C60} (1999) 014903},
\href{http://arxiv.org/abs/hep-ph/9902259}{{\ttfamily arXiv:hep-ph/9902259
  [hep-ph]}}.

\bibitem{Klein:2016yzr}
S.~R. Klein, J.~Nystrand, J.~Seger, Y.~Gorbunov, and J.~Butterworth,
  ``{STARlight: A Monte Carlo simulation program for ultra-peripheral
  collisions of relativistic ions}'',
  \href{http://dx.doi.org/10.1016/j.cpc.2016.10.016}{{\em Comput. Phys.
  Commun.} {\bfseries 212} (2017) 258--268},
\href{http://arxiv.org/abs/1607.03838}{{\ttfamily arXiv:1607.03838 [hep-ph]}}.

\bibitem{Broz:2019kpl}
M.~Broz, J.~G. Contreras, and J.~D. Tapia~Takaki, ``{A generator of forward
  neutrons for ultra-peripheral collisions: $\textbf{n$\mathbf{_O^O}$n}$}'',
  \href{http://dx.doi.org/https://doi.org/10.1016/j.cpc.2020.107181}{{\em
  Comput. Phys. Commun.} (2020) 107181},
\href{http://arxiv.org/abs/1908.08263}{{\ttfamily arXiv:1908.08263 [nucl-th]}}.

\bibitem{Citron:2018lsq}
Z.~Citron {\em et~al.}, ``{Report from Working Group 5}: {Future physics
  opportunities for high-density QCD at the LHC with heavy-ion and proton
  beams}'', \href{http://dx.doi.org/10.23731/CYRM-2019-007.1159}{{\em CERN
  Yellow Rep. Monogr.} {\bfseries 7} (2019) 1159--1410},
  \href{http://arxiv.org/abs/1812.06772}{{\ttfamily arXiv:1812.06772
  [hep-ph]}}.

\bibitem{Abelev:2014ffa}
{\bfseries ALICE} Collaboration, B.~Abelev {\em et~al.}, ``{Performance of the
  ALICE Experiment at the CERN LHC}'',
  \href{http://dx.doi.org/10.1142/S0217751X14300440}{{\em Int. J. Mod. Phys.}
  {\bfseries A29} (2014) 1430044},
\href{http://arxiv.org/abs/1402.4476}{{\ttfamily arXiv:1402.4476 [nucl-ex]}}.

\bibitem{Aamodt:2010aa}
{\bfseries ALICE} Collaboration, K.~Aamodt {\em et~al.}, ``{Alignment of the
  ALICE Inner Tracking System with cosmic-ray tracks}'',
  \href{http://dx.doi.org/10.1088/1748-0221/5/03/P03003}{{\em JINST} {\bfseries
  5} (2010) P03003},
\href{http://arxiv.org/abs/1001.0502}{{\ttfamily arXiv:1001.0502
  [physics.ins-det]}}.

\bibitem{Alme:2010ke}
J.~Alme {\em et~al.}, ``{The ALICE TPC, a large 3-dimensional tracking device
  with fast readout for ultra-high multiplicity events}'',
  \href{http://dx.doi.org/10.1016/j.nima.2010.04.042}{{\em Nucl. Instrum.
  Meth.} {\bfseries A622} (2010) 316--367},
\href{http://arxiv.org/abs/1001.1950}{{\ttfamily arXiv:1001.1950
  [physics.ins-det]}}.

\bibitem{Akindinov:2009zzc}
A.~Akindinov {\em et~al.}, ``{A topological trigger based on the Time-of-Flight
  detector for the ALICE experiment}'',
  \href{http://dx.doi.org/10.1016/j.nima.2008.12.016}{{\em Nucl. Instrum. Meth.
  A} {\bfseries 602} (2009) 372--376}.

\bibitem{Abbas:2013taa}
{\bfseries ALICE} Collaboration, E.~Abbas {\em et~al.}, ``{Performance of the
  ALICE VZERO system}'',
  \href{http://dx.doi.org/10.1088/1748-0221/8/10/P10016}{{\em JINST} {\bfseries
  8} (2013) P10016},
\href{http://arxiv.org/abs/1306.3130}{{\ttfamily arXiv:1306.3130 [nucl-ex]}}.

\bibitem{Loizides:2017ack}
C.~Loizides, J.~Kamin, and D.~d'Enterria, ``{Improved Monte Carlo Glauber
  predictions at present and future nuclear colliders}'',
  \href{http://dx.doi.org/10.1103/PhysRevC.97.054910}{{\em Phys.\ Rev.\ C}
  {\bfseries 97} (2018) 054910},
  \href{http://arxiv.org/abs/1710.07098}{{\ttfamily arXiv:1710.07098
  [nucl-ex]}}. [Erratum: Phys.Rev.C 99, 019901 (2019)].

\bibitem{Acharya:2018hhy}
{\bfseries ALICE} Collaboration, S.~Acharya {\em et~al.}, ``{Centrality and
  pseudorapidity dependence of the charged-particle multiplicity density in
  Xe\textendash{}Xe collisions at $\sqrt{s_{\rm NN}}$ =5.44TeV}'',
  \href{http://dx.doi.org/10.1016/j.physletb.2018.12.048}{{\em Phys. Lett. B}
  {\bfseries 790} (2019) 35--48},
  \href{http://arxiv.org/abs/1805.04432}{{\ttfamily arXiv:1805.04432
  [nucl-ex]}}.

\bibitem{Abelev:2013qoq}
{\bfseries ALICE} Collaboration, B.~Abelev {\em et~al.}, ``{Centrality
  determination of Pb-Pb collisions at $\sqrt{s_{NN}}$ = 2.76 TeV with
  ALICE}'', \href{http://dx.doi.org/10.1103/PhysRevC.88.044909}{{\em Phys. Rev.
  C} {\bfseries 88} no.~4, (2013) 044909},
  \href{http://arxiv.org/abs/1301.4361}{{\ttfamily arXiv:1301.4361 [nucl-ex]}}.

\bibitem{Adam_2016}
{\bfseries ALICE} Collaboration, J.~Adam {\em et~al.}, ``{Centrality Dependence
  of the Charged-Particle Multiplicity Density at Midrapidity in Pb-Pb
  Collisions at $\sqrt{s_{\rm NN}}$ =5.02TeV}'', {\em ALICE Public Notes}
  (2015) . \url{https://cds.cern.ch/record/2118084/}.

\bibitem{Soding:1965nh}
P.~S{\"o}ding, ``{On the Apparent shift of the rho meson mass in
  photoproduction}'',
\href{http://dx.doi.org/10.1016/0031-9163(66)90451-3}{{\em Phys. Lett.}
  {\bfseries 19} (1966) 702--704}.

\bibitem{Jackson:1964zd}
J.~D. Jackson, ``{Remarks on the phenomenological analysis of resonances}'',
\href{http://dx.doi.org/10.1007/BF02750563}{{\em Nuovo Cim.} {\bfseries 34}
  (1964) 1644--1666}.

\bibitem{PDG:2020}
P.~A. Zyla and others (Particle Data~Group), ``{Review of Particle Physics}'',
  \href{http://dx.doi.org/10.1093/ptep/ptaa104}{{\em Prog. Theor. Exp. Phys.}
  {\bfseries 2020} no.~8, (08, 2020) }.
  \url{https://doi.org/10.1093/ptep/ptaa104}. 083C01.

\bibitem{Ross:1965qa}
M.~H. Ross and L.~Stodolsky, ``{Photon dissociation model for vector meson
  photoproduction}'',
\href{http://dx.doi.org/10.1103/PhysRev.149.1172}{{\em Phys. Rev.} {\bfseries
  149} (1966) 1172--1181}.

\bibitem{Kryshen:2019jnz}
{\bfseries ALICE} Collaboration, E.~Kryshen, ``{Overview of ALICE results on
  ultra-peripheral collisions}'',
\href{http://dx.doi.org/10.1051/epjconf/201920401011}{{\em EPJ Web Conf.}
  {\bfseries 204} (2019) 01011}.

\bibitem{Pshenichnov:1999hw}
I.~Pshenichnov, I.~Mishustin, J.~Bondorf, A.~Botvina, and A.~Ilinov,
  ``{Particle emission following Coulomb excitation in ultrarelativistic heavy
  ion collisions}'', \href{http://dx.doi.org/10.1103/PhysRevC.60.044901}{{\em
  Phys. Rev. C} {\bfseries 60} (1999) 044901},
  \href{http://arxiv.org/abs/nucl-th/9901061}{{\ttfamily
  arXiv:nucl-th/9901061}}.

\bibitem{Guzey:2018bay}
V.~Guzey, E.~Kryshen, and M.~Zhalov, ``{Photoproduction of light vector mesons
  in Xe--Xe ultraperipheral collisions at the LHC and the nuclear density of
  Xe-129}'', \href{http://dx.doi.org/10.1016/j.physletb.2018.05.058}{{\em Phys.
  Lett. B} {\bfseries 782} (2018) 251--255},
  \href{http://arxiv.org/abs/1803.07638}{{\ttfamily arXiv:1803.07638
  [hep-ph]}}.

\bibitem{Goncalves:2017wgg}
V.~P. Gon{\c c}alves, M.~V.~T. Machado, B.~Moreira, F.~S. Navarra, and G.~S.
  dos Santos, ``{Color dipole predictions for the exclusive vector meson
  photoproduction in pp , pPb , and PbPb collisions at run 2 LHC energies}'',
  \href{http://dx.doi.org/10.1103/PhysRevD.96.094027}{{\em Phys. Rev. D}
  {\bfseries 96} no.~9, (2017) 094027},
  \href{http://arxiv.org/abs/1710.10070}{{\ttfamily arXiv:1710.10070
  [hep-ph]}}.

\bibitem{Iancu:2003ge}
E.~Iancu, K.~Itakura, and S.~Munier, ``{Saturation and BFKL dynamics in the
  HERA data at small x}'',
  \href{http://dx.doi.org/10.1016/j.physletb.2004.02.040}{{\em Phys. Lett.}
  {\bfseries B590} (2004) 199--208},
\href{http://arxiv.org/abs/hep-ph/0310338}{{\ttfamily arXiv:hep-ph/0310338
  [hep-ph]}}.

\bibitem{Cepila:2016uku}
J.~Cepila, J.~G. Contreras, and J.~D. Tapia~Takaki, ``{Energy dependence of
  dissociative $\mathrm{J/}\psi$ photoproduction as a signature of gluon
  saturation at the LHC}'',
  \href{http://dx.doi.org/10.1016/j.physletb.2016.12.063}{{\em Phys. Lett.}
  {\bfseries B766} (2017) 186--191},
\href{http://arxiv.org/abs/1608.07559}{{\ttfamily arXiv:1608.07559 [hep-ph]}}.

\bibitem{Cepila:2018zky}
J.~Cepila, J.~G. Contreras, M.~Krelina, and J.~D. Tapia~Takaki, ``{Mass
  dependence of vector meson photoproduction off protons and nuclei within the
  energy-dependent hot-spot model}'',
  \href{http://dx.doi.org/10.1016/j.nuclphysb.2018.07.010}{{\em Nucl. Phys. B}
  {\bfseries 934} (2018) 330--340},
  \href{http://arxiv.org/abs/1804.05508}{{\ttfamily arXiv:1804.05508
  [hep-ph]}}.

\bibitem{Armesto:2002ny}
N.~Armesto, ``{A Simple model for nuclear structure functions at small $x$ in
  the dipole picture}'',
  \href{http://dx.doi.org/10.1007/s10052-002-1021-z}{{\em Eur. Phys. J.}
  {\bfseries C26} (2002) 35--43},
\href{http://arxiv.org/abs/hep-ph/0206017}{{\ttfamily arXiv:hep-ph/0206017
  [hep-ph]}}.

\bibitem{Sirunyan:2019nog}
{\bfseries CMS} Collaboration, A.~M. Sirunyan {\em et~al.}, ``{Measurement of
  exclusive $\rho(770)^0$ photoproduction in ultraperipheral pPb collisions at
  $\sqrt{s_\mathrm{NN}} =$ 5.02 TeV}'',
  \href{http://dx.doi.org/10.1140/epjc/s10052-019-7202-9}{{\em Eur. Phys. J. C}
  {\bfseries 79} no.~8, (2019) 702},
  \href{http://arxiv.org/abs/1902.01339}{{\ttfamily arXiv:1902.01339
  [hep-ex]}}.

\bibitem{Breitweg:1997ed}
{\bfseries ZEUS} Collaboration, J.~Breitweg {\em et~al.}, ``{Elastic and proton
  dissociative $\rho^0$ photoproduction at HERA}'',
  \href{http://dx.doi.org/10.1007/s100520050136}{{\em Eur. Phys. J. C}
  {\bfseries 2} (1998) 247--267},
  \href{http://arxiv.org/abs/hep-ex/9712020}{{\ttfamily arXiv:hep-ex/9712020}}.

\bibitem{H1:2020lzc}
{\bfseries H1} Collaboration, V.~Andreev {\em et~al.}, ``{Measurement of
  Exclusive $\pi^{+}\pi^{-}$ and $\rho^0$ Meson Photoproduction at HERA}'',
  \href{http://arxiv.org/abs/2005.14471}{{\ttfamily arXiv:2005.14471
  [hep-ex]}}.

\bibitem{Guzey:2016piu}
V.~Guzey, E.~Kryshen, and M.~Zhalov, ``{Coherent photoproduction of vector
  mesons in ultraperipheral heavy ion collisions: Update for run 2 at the CERN
  Large Hadron Collider}'',
  \href{http://dx.doi.org/10.1103/PhysRevC.93.055206}{{\em Phys. Rev. C}
  {\bfseries 93} no.~5, (2016) 055206},
  \href{http://arxiv.org/abs/1602.01456}{{\ttfamily arXiv:1602.01456
  [nucl-th]}}.

\bibitem{Krelina:2019gee}
M.~Krelina, V.~Gon{\c c}alves, and J.~Cepila, ``{Coherent and incoherent vector
  meson electroproduction in the future electron-ion colliders: the hot-spot
  predictions}'', \href{http://dx.doi.org/10.1016/j.nuclphysa.2019.06.009}{{\em
  Nucl. Phys. A} {\bfseries 989} (2019) 187--200},
  \href{http://arxiv.org/abs/1905.06759}{{\ttfamily arXiv:1905.06759
  [hep-ph]}}.

\end{thebibliography}\endgroup

\newpage
\appendix

%
%

\section{The ALICE Collaboration}
\label{app:collab}
\small
\begin{flushleft} 

S.~Acharya$^{\rm 142}$, 
D.~Adamov\'{a}$^{\rm 97}$, 
A.~Adler$^{\rm 75}$, 
J.~Adolfsson$^{\rm 82}$, 
G.~Aglieri Rinella$^{\rm 35}$, 
M.~Agnello$^{\rm 31}$, 
N.~Agrawal$^{\rm 55}$, 
Z.~Ahammed$^{\rm 142}$, 
S.~Ahmad$^{\rm 16}$, 
S.U.~Ahn$^{\rm 77}$
% ...

\bigskip

\bigskip 

\textbf{\Large Affiliation Notes}

\bigskip 

% ...
$^{\rm V}$ Also at: Institute of Theoretical Physics, University of Wroclaw, Poland\\

\bigskip

\bigskip 

\textbf{\Large Collaboration Institutes}

\bigskip 

% ...
$^{16}$ Department of Physics, Aligarh Muslim University, Aligarh, India\\
% ...
$^{31}$ Dipartimento DISAT del Politecnico and Sezione INFN, Turin, Italy\\
% ...
$^{35}$ European Organization for Nuclear Research (CERN), Geneva, Switzerland\\
% ...
$^{55}$ INFN, Sezione di Bologna, Bologna, Italy\\
% ...
$^{75}$ Johann-Wolfgang-Goethe Universit\"{a}t Frankfurt Institut f\"{u}r Informatik, Fachbereich Informatik und Mathematik, Frankfurt, Germany\\
% ...
$^{77}$ Korea Institute of Science and Technology Information, Daejeon, Republic of Korea\\
% ...
$^{82}$ Lund University Department of Physics, Division of Particle Physics, Lund, Sweden\\
% ...
$^{97}$ Nuclear Physics Institute of the Czech Academy of Sciences, \v{R}e\v{z} u Prahy, Czech Republic\\
% ...
$^{142}$ Variable Energy Cyclotron Centre, Homi Bhabha National Institute, Kolkata, India\\
% ...

\end{flushleft} 

\end{document}